\documentclass[prl,reprint,nofootinbib,altaffilletter,superscriptaddress,showpacs]{revtex4-1}
\usepackage{amsmath,amssymb,amsfonts}
\usepackage[dvips]{graphicx}
\usepackage{dcolumn}
\usepackage{bm}
\usepackage{bbm}
\usepackage{xcolor}
\usepackage{hyperref}
\usepackage{lineno}
\usepackage{ulem}

\newcommand{\bibprl}[3]{Phys.\ Rev.\ Lett.\ {\bf #1}, #2 (#3)}
\newcommand{\bibprc}[3]{Phys.\ Rev.\ C {\bf #1}, #2 (#3)}
\newcommand{\bibprd}[3]{Phys.\ Rev.\ D {\bf #1}, #2 (#3)}
\newcommand{\bibplb}[3]{Phys.\ Lett.\ B {\bf #1}, #2 (#3)}
\newcommand{\bibnpa}[3]{Nucl.\ Phys.\ A {\bf #1}, #2 (#3)}
\newcommand{\bibnima}[3]{Nucl.\ Instrum.\ Meth.\ A {\bf #1}, #2 (#3)}
\newcommand{\bibepja}[3]{Eur.\ Phys.\ J.\ A {\bf #1}, #2 (#3)}

\def\figwidth{0.475\textwidth}
\begin{document}

\title{Resonance-like structure near the $\eta d$ threshold in
the $\gamma{d}${$\to$}$\pi^0\eta{d}$ reaction}
\author{T.~Ishikawa}
\email[Corresponding author: ]{ishikawa@lns.tohoku.ac.jp}
\affiliation{Research Center for Electron Photon Science (ELPH), Tohoku University, Sendai 982-0826, Japan}
\author{H.~Fujimura}
\altaffiliation[Present address: ]{Department of Physics, Wakayama Medical University, Wakayama 641-8509, Japan}
\affiliation{Research Center for Electron Photon Science (ELPH), 
Tohoku University, Sendai 982-0826, Japan}
\author{H.~Fukasawa}
\affiliation{Research Center for Electron Photon Science (ELPH), 
Tohoku University, Sendai 982-0826, Japan}
\author{R.~Hashimoto}
\altaffiliation[Present address: ]{Institute of Materials Structure Science (IMSS), KEK, Tsukuba 305-0801, Japan}
\affiliation{Research Center for Electron Photon Science (ELPH), 
Tohoku University, Sendai 982-0826, Japan}
\author{Q.~He}
\altaffiliation[Present address: ]{Department of Nuclear Science and Engineering, Nanjing University of Aeronautics and Astronautics (NUAA), Nanjing 210016, China}
\affiliation{Research Center for Electron Photon Science (ELPH), 
Tohoku University, Sendai 982-0826, Japan}
\author{Y.~Honda}
\affiliation{Research Center for Electron Photon Science (ELPH), 
Tohoku University, Sendai 982-0826, Japan}
\author{T.~Iwata}
\affiliation{Department of Physics, Yamagata University, Yamagata 990-8560, Japan}
\author{S.~Kaida}
\affiliation{Research Center for Electron Photon Science (ELPH), 
Tohoku University, Sendai 982-0826, Japan}
\author{J.~Kasagi}
\affiliation{Research Center for Electron Photon Science (ELPH), 
Tohoku University, Sendai 982-0826, Japan}
\author{A.~Kawano}
\affiliation{Department of Information Science, Tohoku Gakuin University, Sendai 981-3193, Japan}
\author{S.~Kuwasaki}
\affiliation{Research Center for Electron Photon Science (ELPH), 
Tohoku University, Sendai 982-0826, Japan}
\author{K.~Maeda}
\affiliation{Department of Physics, Tohoku University, Sendai 980-8578, Japan}
\author{S.~Masumoto}
\affiliation{Department of Physics, University of Tokyo, Tokyo 113-0033, Japan}
\author{M.~Miyabe}
\affiliation{Research Center for Electron Photon Science (ELPH), 
Tohoku University, Sendai 982-0826, Japan}
\author{F.~Miyahara}
\altaffiliation[Present address: ]{Accelerator Laboratory, KEK, Tsukuba 305-0801, Japan}
\affiliation{Research Center for Electron Photon Science (ELPH), 
Tohoku University, Sendai 982-0826, Japan}
\author{K.~Mochizuki}
\affiliation{Research Center for Electron Photon Science (ELPH), 
Tohoku University, Sendai 982-0826, Japan}
\author{N.~Muramatsu}
\affiliation{Research Center for Electron Photon Science (ELPH), 
Tohoku University, Sendai 982-0826, Japan}
\author{A.~Nakamura}
\affiliation{Research Center for Electron Photon Science (ELPH), 
Tohoku University, Sendai 982-0826, Japan}
\author{K.~Nawa}
\affiliation{Research Center for Electron Photon Science (ELPH), 
Tohoku University, Sendai 982-0826, Japan}
\author{Y.~Obara}
\affiliation{Department of Physics, University of Tokyo, Tokyo 113-0033, Japan}
\author{S.~Ogushi}
\affiliation{Research Center for Electron Photon Science (ELPH), 
Tohoku University, Sendai 982-0826, Japan}
\author{Y.~Okada}
\affiliation{Research Center for Electron Photon Science (ELPH), 
Tohoku University, Sendai 982-0826, Japan}
\author{K.~Okamura}
\affiliation{Research Center for Electron Photon Science (ELPH), 
Tohoku University, Sendai 982-0826, Japan}
\author{Y.~Onodera}
\affiliation{Research Center for Electron Photon Science (ELPH), 
Tohoku University, Sendai 982-0826, Japan}
\author{K.~Ozawa}
\affiliation{Institute of Particle and Nuclear Studies (IPNS), High Energy Accelerator Research Organization (KEK), Tsukuba 305-0801, Japan}
\author{Y.~Sakamoto}
\affiliation{Department of Information Science, Tohoku Gakuin University, Sendai 981-3193, Japan}
\author{M.~Sato}
\affiliation{Research Center for Electron Photon Science (ELPH), 
Tohoku University, Sendai 982-0826, Japan}
\author{H.~Shimizu}
\affiliation{Research Center for Electron Photon Science (ELPH), 
Tohoku University, Sendai 982-0826, Japan}
\author{H.~Sugai}
\altaffiliation[Present address: ]{Gunma University Initiative for Advanced Research (GIAR), Maebashi 371-8511, Japan}
\affiliation{Research Center for Electron Photon Science (ELPH), 
Tohoku University, Sendai 982-0826, Japan}
\author{K.~Suzuki}
\altaffiliation[Present address: ]{The Wakasa Wan Energy Research Center, Tsuruga 914-0192, Japan}
\affiliation{Research Center for Electron Photon Science (ELPH), 
Tohoku University, Sendai 982-0826, Japan}
\author{Y.~Tajima}
\affiliation{Department of Physics, Yamagata University, Yamagata 990-8560, Japan}
\author{S.~Takahashi}
\affiliation{Research Center for Electron Photon Science (ELPH), 
Tohoku University, Sendai 982-0826, Japan}
\author{Y.~Taniguchi}
\affiliation{Research Center for Electron Photon Science (ELPH), 
Tohoku University, Sendai 982-0826, Japan}
\author{Y.~Tsuchikawa}
\altaffiliation[Present address: ]{J-PARC Center, Japan Atomic Energy Agency (JAEA), Tokai 319-1195, Japan}
\affiliation{Research Center for Electron Photon Science (ELPH), 
Tohoku University, Sendai 982-0826, Japan}
\author{H.~Yamazaki}
\altaffiliation[Present address: ]{Radiation Science Center, KEK, Tokai 319-1195, Japan}
\affiliation{Research Center for Electron Photon Science (ELPH), 
Tohoku University, Sendai 982-0826, Japan}
\author{R.~Yamazaki}
\affiliation{Research Center for Electron Photon Science (ELPH), 
Tohoku University, Sendai 982-0826, Japan}
\author{H.Y.~Yoshida}
\affiliation{Department of Physics, Yamagata University, Yamagata 990-8560, Japan}

\begin{abstract}
To investigate the interaction between the nucleon $N$ and 
nucleon resonance
$N(1535)1/2^-$, the 
$\eta d$ threshold structure
connected to the isoscalar $S$-wave $N$-$N(1535)1/2^-$ system 
has been
experimentally studied 
in the $\gamma{d}${$\to$}$\pi^0\eta{d}$ reaction 
at incident photon energies ranging from the reaction threshold to 1.15 GeV.
A strong enhancement is observed near the $\eta d$ threshold over the 
three-body phase-space contribution in the $\eta d$ 
invariant-mass
 distribution. 
An analysis incorporating the known isovector resonance $\mathcal{D}_{12}$ with 
a 
spin-parity
of $2^+$ in the $\pi^0d$ channel 
reveals
the existence of a 
narrow isoscalar resonance-like structure with $1^-$ in the $\eta d$ system.
Using a Flatt\'e parametrization,
the mass is found to be $2.427_{-0.006}^{+0.013}$ GeV, 
close to the $\eta d$ threshold,
and the width is
$\left(0.029_{-0.029}^{+0.006}{\rm\ GeV}\right)
+
\left(0.00_{-0.00}^{+0.41}\right) p_\eta c$,
where $p_\eta$ denotes the $\eta$ momentum in the rest frame of 
the $\eta d$ system.
The observed structure would be attributed to
a predicted isoscalar $1^-$ $\eta NN$ bound state from $\eta NN$ 
and $\pi NN$ coupled-channel calculation, or an $\eta d$ virtual state
owing to strong $\eta d$ attraction.
\end{abstract}


\maketitle
The structure 
and interaction of hadrons provides crucial insight into the non-perturbative mechanisms in quantum 
chromodynamics.
Attraction between an $\eta$ meson and 
a nucleon $N$ makes it possible to form
 an $\eta$-mesic nucleus, as predicted by 
Haider and Liu~\cite{haider-liu}.
This is an exotic state in which $\eta$ is bound 
to the nucleus by the strong 
interaction force alone, and allows the study of
the behavior of $\eta$ in a dense nuclear environment. 
The binding energy of $\eta$ in the nuclear medium is sensitive 
to a singlet component of $\eta$ ($\eta$-$\eta'$ mixing)~\cite{mix1,mix2,mix3}.
The level and width of an $\eta$-mesic nucleus can yield the in-medium properties 
of nucleon resonance $N(1535)1/2^-$ ($N^*$)~\cite{jido1,jido2,jido3,jido4}, 
which is speculated to be the chiral partner of $N$.
This is because $\eta$ in the nuclear medium is expressed 
by mixing of the $\eta$-mesonic state, and 
the $N^*$-particle and $N$-hole excitation state. 

Exotic $\eta$-mesic nuclei have been intensively investigated theoretically,
and searched for experimentally~\cite{rev1,rev2,rev3}.
Experimental hints of possible $\eta$-mesic nuclei 
have been obtained in the threshold behavior of $\eta$-production reactions.
The existence of an $\eta$-mesic nucleus enhances
the cross section near the reaction threshold 
compared with phase space.
The total cross section shows a steep increase 
from the threshold 
in $\eta\,{}^3{\rm He}$ production from the $pd$ collisions~\cite{3he0,3he1,3he2}.
The
possibility of an 
$\eta\,{}^3{\rm He}$ weakly bound state is claimed
by analyzing the $\eta$ angular distribution~\cite{3he3}.
If an $\eta\,{}^3{\rm He}$ bound state exists,
it should appear independently of the initial state of reactions.
Coherent $\eta$ photoproduction on $ {}^3{\rm He}$ also shows 
a strong threshold enhancement,
and the angular distribution of $\eta$ emission is rather flat 
near the threshold as compared with the expected distribution
based on the nuclear form factor~\cite{3he4}.
The WASA-at-COSY collaboration has searched for $\eta$-mesic $^3$He and $^4$He nuclei 
in the $pd$ and $dd$ reactions, respectively, by detecting daughter particles from $\eta$ or $N^*$ in a 
nucleus~\cite{wasa2013,wasa2017,wasa2018,wasa2020a,wasa2020b}.
No convincing evidence for an $\eta$-mesic nucleus has yet been obtained.

An $\eta d$ bound state, if it exists,
is the lightest $\eta$-mesic nucleus. 
The $S$-wave $\eta d$ system has an isospin of 0 and a 
spin-parity
$J^\pi$ of $1^-$, and its properties are connected to the $NN^*$ interaction.
An $\eta NN$ bound state has been predicted near the $\eta d$ threshold with a width $\Gamma$ of 0.01--0.02 GeV~\cite{etad1}
from the three-body $\eta NN$-$\pi NN$ coupled-channel calculation. 
This state can be located lower than the threshold by 8 MeV~\cite{ueda92}.
This state is suggested by the
significant deviation from phase space near the threshold 
in $\eta d$ production from the $pn$ 
collisions~\cite{etad3,etad4,cale97,cale98,bilg04}.
In contrast, the $\gamma d\to \eta d$ reaction
does not show any indication of this state;
its angular distribution is explained by 
the quasi-free (QF) $\gamma N\to \eta N$ process~\cite{hoff97,weis01}. 
The possibility of an $\eta d$ bound state is ruled out in several
theoretical three-body calculations for various $\eta N$ scattering 
parameters~\cite{delo00,garc00,barn15}.
Instead, a narrow $\eta d$ virtual state is 
inferred to reproduce the experimental data~\cite{wyce01,fix02,garc03}.
However, qualitative disagreement is still observed in different theoretical 
calculations which cannot be explained by uncertainties in
the $\eta N$ scattering parameters alone.
It should be noted that the existence of an $\eta d$ bound or resonance
state near the threshold is claimed in Ref.~\cite{shev00}.

The $\gamma d \to\pi^0\eta d$ reaction can provide a condition 
of low $\eta d$ relative momentum,
producing a possible $\eta d$ bound state.
In a similar $\gamma d\to \pi^0\pi^0d$ reaction,
a sequential process
$\gamma d\to \mathcal{D}_{\rm IS} \to \pi^0 \mathcal{D}_{\rm IV}\to \pi^0\pi^0 d$
~\cite{plb2} is dominant,
where $\mathcal{D}_{\rm IS}$ and $\mathcal{D}_{\rm IV}$
denote isoscalar and isovector states with a baryon number of 2, respectively.
In $\gamma d \to\pi^0\eta d$,
the two sequential processes,
$\gamma d \to \mathcal{D}_{\rm IV} \to \pi^0 \mathcal{D}_{\rm IS}\to \pi^0\eta d$ and
$\gamma d \to \mathcal{D}_{\rm IV} \to \eta \mathcal{D}_{\rm IV}'\to \pi^0\eta d$,
are expected to be observed, and the tail of the possible $\eta d$ bound state
appears as $\mathcal{D}_{\rm IS}$.
In this letter, we study the $\gamma d\to \pi^0\eta d$
reaction
to clarify the structure that appears in the low-relative-momentum region
of the $\eta d$ system generated after $\pi^0$ emission.

A series of experiments~\cite{exp}
were carried out using a bremsstrahlung photon beam~\cite{tag2,bpm,trans,wire} from 
1.20-GeV electrons circulating
in a synchrotron~\cite{stb} at the Research Center for Electron Photon
Science (ELPH), Tohoku University, Japan~\cite{elph}.
The photon energy ranging from 0.75 to 1.15 GeV
was determined by detecting 
the post-bremsstrahlung electron with a 
photon-tagging counter, STB-Tagger II~\cite{tag2}.
The target used was liquid deuterium with a thickness of 45.9~mm.
All the final-state particles in the $\gamma{d} \to \pi^0\eta d\to\gamma\gamma\gamma\gamma d$ reaction
were measured with the FOREST detector consisting of 
three different electromagnetic calorimeters (EMCs)~\cite{forest}.
A plastic-scintillator hodoscope (PSH) was placed in front of each EMC
to identify charged particles.
The forward PSH could determine their impact positions. 
The trigger condition of the data acquisition
required detection of multiple particles in coincidence with a 
photon-tagging signal~\cite{plb2,forest,plb1,prc}.

Initially, events containing four neutral particles and a charged particle 
were selected. 
An EMC cluster without a corresponding PSH hit was recognized as a neutral particle.
A PSH hit gave a charged particle regardless of the existence of a corresponding EMC cluster.
The time difference between every two neutral clusters
out of four was required to be less than thrice that
of the time resolution.
The selected events were those in which the charged particle was detected with the forward PSH,
under the conditions that 
the time delay from the response of the four neutral clusters was longer than 1 ns,
and the deposit energy of a charged particle in PSH was greater than twice 
that of the minimum ionizing particle.
Further selection was made by applying a kinematic fit (KF) with six constraints:
 energy and three-momentum conservation,
the invariant mass of two photons out of four being the $\pi^0$ mass,
and that of the other two being the $\eta$ mass.
The most probable combination was selected in each event for 
$\pi^0\to\gamma\gamma$ and $\eta\to\gamma\gamma$.
The momentum of the charged particle was obtained from the time delay
assuming that the charged particle had the deuteron mass.
Events with $\chi^2$ probability higher than 0.2 
were selected to discriminate from other background reactions.
The most competitive background was 
from deuteron misidentification 
events in the QF $\gamma{p'}${$\to$}$\pi^0\eta p$ reaction.
Thus, selected events were additionally required 
to exhibit $\chi^2$ probability below 0.01 
in another KF for the $\gamma{p'}${$\to$}$\pi^0\eta p$ hypothesis
where the $x$, $y$, and $z$ momenta of the initial bound proton were 
assumed to be measured with a centroid of 0 MeV/$c$ and
a resolution of 40 MeV/$c$,
and the total energy of the bound proton was given 
assuming the on-shell spectator neutron.
Finally, sideband-background subtraction was performed
for accidental-coincidence events detected in STB-Tagger II and FOREST.

\begin{figure}[b]
\begin{center}
\includegraphics[width=\figwidth]{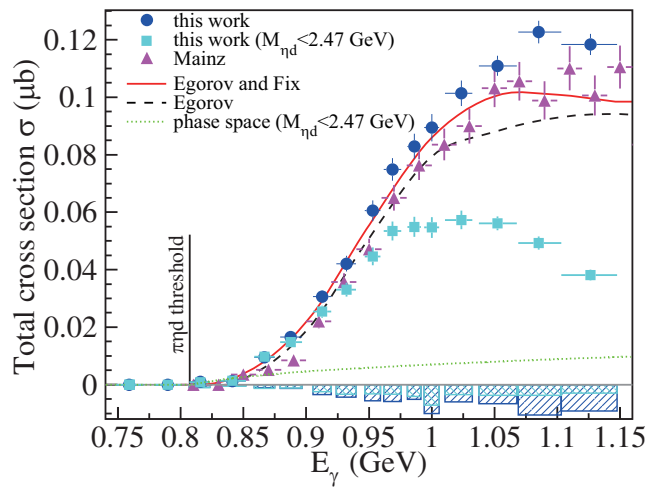}
\end{center}
\caption{Total cross section $\sigma$ as a function of the incident photon energy $E_\gamma$.
The circles (blue) show $\sigma$ obtained in this work, and the triangles (magenta) 
show that obtained at the Mainz MAMI facility~\cite{mainz}.
The horizontal error of each data point corresponds to 
the $E_\gamma$ coverage, and the vertical error 
corresponds to the statistical error of $\sigma$.
The solid (red) and dashed (black) curves show theoretical calculations 
with $\eta d$ and other FSIs in Ref.~\cite{fix2} and 
Ref.~\cite{ego}, respectively.
The squares (cyan) show $\sigma$ obtained for the
events with $M_{\eta d}<2.47$ GeV,
and the dotted curve (green) shows the corresponding phase space contribution.
The lower hatched histograms show the systematic errors of $\sigma$
with right-up straight lines for all the events (blue) and 
with left-up lines for $M_{\eta d}<2.47$ GeV (cyan).
}
\label{fig1}
\end{figure}
The total cross section was obtained 
by estimating the acceptance of
$\gamma\gamma\gamma\gamma{d}$ detection in 
a Monte Carlo simulation based on Geant4~\cite{geant4a,geant4b,geant4c}.
Here,
 event generation was modified from pure phase space
to reproduce the following three measured distributions:
the $\pi^0{d}$ invariant mass $M_{\pi{d}}$, 
the $\eta{d}$ invariant mass $M_{\eta{d}}$, 
and 
the deuteron emission angle $\cos\theta_d$ in the 
$\gamma{d}$ center-of-mass (CM)
frame.
Figure~\ref{fig1} shows the total cross section $\sigma$ as a function of the incident photon energy $E_\gamma$ (excitation function).
The data obtained in this work
were consistent with those obtained at the Mainz MAMI facility~\cite{mainz}.
The systematic uncertainty of $\sigma$ is also given in Fig.~\ref{fig1}. 
It includes the uncertainty of event selection in KF;
that of acceptance owing to the uncertainties in the
$M_{\pi{d}}$, $M_{\eta{d}}$, and $\cos\theta_d$ 
distributions in event generation of the simulation;
that of deuteron detection efficiency;
and that of normalization resulting from the 
numbers of target deuterons and incident photons.
In Fig.~\ref{fig1}, the data are compared with the existing
theoretical calculations with the final-state interactions (FSIs)
including $\eta d$ by Egorov and Fix~\cite{fix2} (red solid)
and by Egorov~\cite{ego} (black dashed).
The excitation function is well-reproduced by these calculations 
near the threshold.

\begin{figure}[b]
\begin{center}
\includegraphics[width=\figwidth]{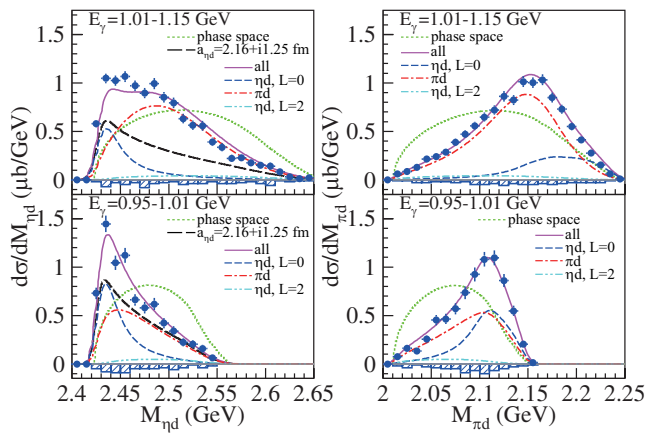}
\end{center}
\caption{Differential cross sections $d\sigma/dM_{\eta d}$ 
(left)
and $d\sigma/dM_{\pi d}$ (right)
at $E_\gamma=1.01$--1.15 GeV (top) and 0.95--1.01 GeV (bottom).
}
\label{fig2}
\end{figure}
To study the $\eta d$ threshold structure,
we obtained the differential cross section $d\sigma/dM_{\eta d}$ 
at $E_\gamma=1.01$--1.15 and 0.95--1.01 GeV
as shown in Fig.~\ref{fig2} (left).
The experimental data are presented by the circles with statistical errors, 
and the systematic uncertainties by the hatched histograms. 
An enhancement is observed in $d\sigma/dM_{\eta d}$
 over the phase-space 
contribution (green dotted) in the low-mass region.
This enhancement is much broader at
$E_\gamma=1.01$--1.15 GeV than 
at $E_\gamma=0.95$--1.01 GeV, 
suggesting the appearance of another contribution 
from a resonance in the $\pi d$ channel.
We also obtained the differential cross section $d\sigma/dM_{\pi d}$ 
similarly to $d\sigma/dM_{\eta d}$ as shown in Fig.~\ref{fig2} (right).
In $d\sigma/dM_{\pi d}$,
we observe a significant enhancement at high masses,
corresponding to the known isovector $\mathcal{D}_{12}$ resonance 
with $J^\pi=2^+$, $M\simeq 2.14$ GeV, and 
$\Gamma\simeq 0.09$ GeV~\cite{plb2}.

Only the $S$-wave $\eta d$ system ($\mathcal{D}_{\eta d}$) forms a peak close to the threshold in $d\sigma/dM_{\eta{d}}$. 
The $\mathcal{D}_{\eta d}$ system with $J^\pi=1^-$ 
decays into $\eta d$ dominantly in the $S$ wave, 
and possibly in the $D$ wave. 
The $S$- and $D$-wave contributions 
to the $d\sigma/dM_{\eta d}$
distribution differ in shape.
A fraction of the $D$-wave contribution to the $S$-wave at a fixed $M_{\eta d}$
is proportional to $p_{\eta}^4$,
where $p_{\eta}$ denotes the $\eta$ momentum 
in the rest frame of the $\eta d$ system.
The $\mathcal{D}_{\eta d}$ and $\mathcal{D}_{12}$
contributions are separated by fitting a set of functions, expressed as the
sum of $S$- and $D$-wave decay contributions of a Breit-Wigner (BW) resonance
in the $\eta d$ channel and $\mathcal{D}_{12}$ contribution in the $\pi^0d$ channel,
to the $M_{\eta d}$ and $M_{\pi d}$ data 
at $E_\gamma=1.01$--1.15 and 0.95--1.01
GeV simultaneously.
The function for $d\sigma/dM_{\eta d}$ is given by 
\begin{equation}
\displaystyle
\frac{d\sigma}{dM_{\eta d}}\!\left(
M_{\eta d}
\right)
=\alpha_0 \int\! A (M_{\eta d},M_{\pi d})\,
V_{\rm PS}(M_{\eta d},M_{\pi d})\,
dM_{\pi d}
\label{eq:fit1}
\end{equation}
where $V_{\rm PS}(M_{\eta d},M_{\pi d})$ expresses the phase-space contribution
and
$A(M_{\eta d},M_{\pi d})$ gives the enhancement owing to
 the two resonances:
\begin{equation}
\displaystyle
A (M_{\eta d},M_{\pi d}) = \left(1+\alpha_2 p_{\eta}^4 \right) L_{M,\Gamma}^{\mathcal{D}_{\eta d}}
\!\left( M_{\eta d} \right) +
\alpha _1 
L_{M,\Gamma}^{\mathcal{D}_{12}}\!\left( M_{\pi d}\right).
\label{eq:fit2}
\end{equation}
Here, $L_{M,\Gamma}^{\mathcal{D}_{\eta d}}\!\left( M_{\eta d} \right)$ 
and $L_{M,\Gamma}^{\mathcal{D}_{12}}\!\left( M_{\pi d} \right)$ 
represent BW distributions with 
$M$ and $\Gamma$ for $\mathcal{D}_{\eta d}$ and $\mathcal{D}_{12}$, respectively.
To incorporate the opening of the $\eta d$ channel, 
the $\eta d$ partial width is parametrized by the effective coupling $g$
(known as the Flatt\'e parametrization~\cite{flat76,kala09}):
\begin{equation}
\Gamma= \Gamma_0 + g p_\eta c
\end{equation} 
for $\mathcal{D}_{\eta d}$,
and $\Gamma_0$ is the width for the other open channels ($NN$, $\pi NN$,
and $\pi \pi NN$).
It should be noted that 
the phase space of the $\eta d$ decay is taken into account 
in $V_{\rm PS}(M_{\eta d},M_{\pi d})$.
The other $M$ and $\Gamma$ parameters 
are assumed to be constant.
Equation~(\ref{eq:fit1})
 is evaluated by the convolution of a Gaussian with an experimental mass resolution 
of $\sigma_{M_{\eta d}}$ = 6.0 (4.8) MeV at $E_\gamma=1.01$--1.15 (0.95--1.01) GeV.
The function for $d\sigma/dM_{\pi d}$ is given by 
\begin{equation}
\displaystyle
\frac{d\sigma}{dM_{\pi d}}\!\left(
M_{\pi d}
\right)
=\alpha_0 \int\! A (M_{\eta d},M_{\pi d})\,
V_{\rm PS}(M_{\eta d},M_{\pi d})\,
dM_{\eta d}
\end{equation}
with $\sigma_{M_{\pi d}}$ = 6.1 (4.8) MeV at $E_\gamma=1.01$--1.15 (0.95--1.01) GeV.
In Fig.~\ref{fig2}, the mass spectrum is an incoherent sum of 
two resonances for the following reason. 
As discussed later, we consider
that $\mathcal{D}_{\eta d}$ and $\mathcal{D}_{12}$ are produced in paths (\ref{eq:seq1}) and
 (\ref{eq:seq2}),  respectively. The mass spectrum is a plot of the integrated yield 
for the angular distributions of $\pi^0$ and $\eta$. 
The interference term of the two paths is 
zero unless $L_1\left(\pi^0\right) = L_2\left(\pi^0\right)$
and $L_2\left(\eta\right)=L_1\left(\eta\right)$, owing to the orthogonality of the spherical harmonics
which appear in the angular component of the wave function.
In the analysis, the $L_1\left(\eta\right) =1$ component is deduced
to be $\sim 100\%$; 
therefore, almost no effect of the interference 
effect
is expected in the mass spectra.

The obtained parameters in the fit 
are 
$\left(M,\Gamma_0, g \right)=\left(2.427^{-0.006}_{+0.013}{\rm\ GeV}, 
0.029_{-0.029}^{+0.006}{\rm\ GeV},
0.00_{-0.00}^{+0.41}\right)$ for $\mathcal{D}_{\eta d}$, and
$\left(M,\Gamma\right)=\left(2.158^{-0.003}_{+0.003}, 0.116_{-0.011}^{+0.005}\right)$ GeV
for $\mathcal{D}_{12}$,
where $\chi^2=131.4$ and the number of data points is 76.
Also plotted in Fig.~\ref{fig2} are 
the $S$- (blue dashed) and $D$-wave
(cyan double-dotted)
 decay contributions as well as the
$\mathcal{D}_{12}$ contribution (red dot-dashed).
Each of $d\sigma/dM_{\eta d}$ and $d\sigma/dM_{\pi d}$ consists of two peaks.
The centroid of the low-mass peak in $d\sigma/dM_{\eta d}$
is close to the $\eta d$ threshold independently of $E_\gamma$.
The high-mass peak reflects the appearance of the 2.14-GeV peak in $d\sigma/dM_{\pi d}$.
The centroid of the high-mass peak decreases with a decrease of
$E_\gamma$. 
Because the $M_{\pi d}$ coverage is limited at $E_\gamma=0.95$--1.01 GeV, 
the two peaks merge
into a bump in $d\sigma/dM_{\pi{d}}$ 
with substantial
distortion of the 2.14-GeV peak.
It is thus revealed that a narrow resonance-like structure exists in the 
vicinity of the $\eta d$ threshold (2.423 GeV).
This is not observed 
due to its isoscalar nature
in $\gamma d\to \eta d$ where the QF process is dominant.

The $S$-wave $\eta d$ resonance states with 
widths broader than 0.05 GeV are ruled out for the threshold enhancement in $d\sigma/dM_{\eta d}$.
It could be attributed to
the predicted $\eta d$ bound state~\cite{etad1,ueda92,shev00},
being a Feshbach resonance in the $\eta d$ and isoscalar $\pi NN$ coupled channels.
If so, the corresponding enhancement can be observed in the isoscalar $\pi N N$ and $\pi^0\pi^0 d$ channels (corresponding to the $\Gamma_0\ne 0$ case).
Possibly related to this is a bump 
observed by the WASA-at-COSY collaboration
at the CM energy of $\sim 2.31$ GeV in the isoscalar $NN\to\pi NN$ reaction~\cite{adla17}.
The 
spin-parity
of this bump is not clear ($1^+$ and $0^+$ are discussed in Ref.~\cite{kuku20,clem20}),
and the bump may include a $1^-$ state~\cite{clem21} 
corresponding to the $\eta d$ bound state.

The threshold enhancement can also be interpreted as
an $\eta d$ virtual state~\cite{wyce01,fix02,garc03} (corresponding to
the $\Gamma_0=0$ case).
The square of  the amplitude would be proportional to  
$\left|a_{\eta d}^{-1}-ip_\eta\right|^{-2}$
for production of an $\eta d$ system at low relative momentum $p_\eta$,
where $a_{\eta d}$ denotes the $\eta d$ scattering length.
Using $a_{\eta d} = 2.16+i1.25$ fm~\cite{fix18}
extracted from $pn\to\eta d$~\cite{cale98,bilg04},
the $M_{\eta d}$ distributions are expected as shown
by the long-dashed curves (black) in Fig.~\ref{fig2}(left),
and similar to the decomposed $\mathcal{D}_{\eta d}$ 
contributions close to the threshold.
They are observed in a wider range as compared with the 10-MeV 
range of $pn\to \eta d$.
High $\eta d$ angular momenta would be suppressed 
in $\gamma d\to\pi^0\eta d$
because sequential processes are dominant.
Additionally, the $pn\to \eta d$ data may be affected by 
FSI between $\eta$ and the spectator proton.

\begin{figure}[b]
\begin{center}
\includegraphics[width=\figwidth]{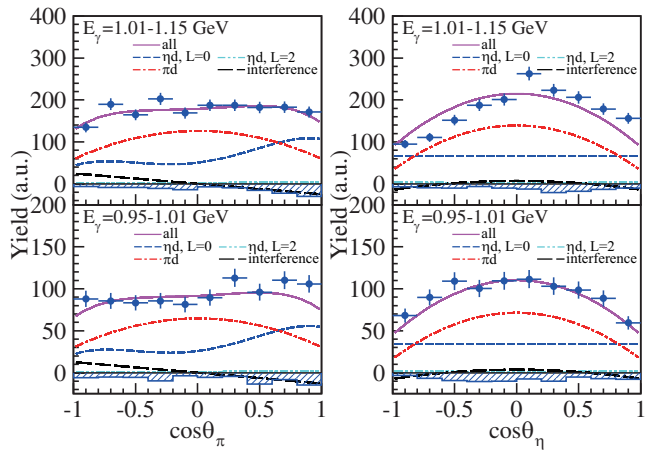}
\end{center}
\caption{
Acceptance-corrected angular distributions for $\pi^0$ 
with respect to the photon beam direction
in the $\gamma d$-CM frame (left)
and for $\eta$ with respect to the opposite direction to $\pi^0$
in the $\eta d$ rest frame (right)
at $E_\gamma= 1.01$--$1.15$ GeV (top) and $0.95$--$1.01$ GeV (bottom).
Events with $M_{\eta d} \le 2.47$ GeV are selected. 
}\label{fig3}
\end{figure}
The $\pi^0\mathcal{D}_{\eta d}$- and $\eta\mathcal{D}_{12}$-produced processes
are investigated using
angular distributions of $\pi^0$ and $\eta$
obtained for 
the events with $M_{\eta{d}}<2.47$ GeV as shown in Fig.~\ref{fig3}.
Figure~\ref{fig3}
 (left) shows the deduced $\pi^0$ angular distributions in
the $\gamma{d}$-CM frame with respect to the incident photon direction;
the experimental distributions are almost flat.
The $\eta$ angular distributions in the $\eta d$ rest frame
with respect to the opposite direction to $\pi^0$ emission
are shown in Fig.~\ref{fig3} (right). 
They take a convex-upward shape, and show almost 
$90^\circ$ symmetry.
Thus, contamination of a state with $J^\pi$ other than
$1^-$ ($2^+$) is assumed to be negligibly small in 
$\mathcal{D}_{\eta d}$ ($\mathcal{D}_{12}$).

We calculated the $\pi^0$ and $\eta$ angular distributions for the reaction sequences of interest 
using the density matrix (statistical tensor) formalism~\cite{rose}:
\begin{widetext}
\begin{equation}
J_0\left(d\right)=1
\xrightarrow{L_0\left(\gamma\right)} 
J_1\left(\pi^0\mathcal{D}_{\eta d}\right) \xrightarrow{L_1\left(\pi^0\right)} J_2\left(\mathcal{D}_{\eta d}\right)\mbox{$=$}1
\xrightarrow{L_2\left(\eta\right)=0, 2}
J_3\left(d\right)=1
\label{eq:seq1}
\end{equation}
and
\begin{equation}
J_0\left(d\right)=1
\xrightarrow{L_0\left(\gamma\right)} 
J_1\left(\eta \mathcal{D}_{12}\right)
\xrightarrow{L_1\left(\eta\right)} 
J_2\left(\mathcal{D}_{12}\right)\mbox{$=$}2
\xrightarrow{L_2\left(\pi^0\right)=1}
J_3\left(d\right)=1,
\label{eq:seq2}
\end{equation}
\end{widetext}
where 
$J_1$ and $J_2$ denote the spins of first and second intermediate states, respectively,
and $J_0=J_3=1$ are those of the initial and final deuteron.
The 
$L_0$ denotes the angular momentum of
 the incident photons.
The $L_1$ and $L_2$ denote the angular momenta of
meson emission from the first and second intermediate states,
respectively.
A set of the
amplitudes $A_{\Lambda\Lambda}$ 
was determined for all the $\Lambda=(L_0,J_1,L_1,J_2,L_2)$ combinations
with $L_0, J_1, L_1 \le 2$ 
to reproduce the measured angular 
distributions 
at $E_\gamma= 1.01$--$1.15$ and $0.95$--$1.01$ GeV
simultaneously (40 data points). 
The amplitude for a mixed state is given
by $A_{\Lambda\Lambda'}${$=$}$\left(A_{\Lambda\Lambda}A_{\Lambda'\Lambda'}\right)^{1/2}$.
The $L_2=2$ amplitudes are given by 
$A_{\Lambda_2\Lambda_2} = A_{\Lambda_0\Lambda_0} \tan\phi$,
where $L_2=0$ in $\Lambda_0$ is replaced by $L_2=2$ in $\Lambda_2$.
The $L_2=1$ amplitudes are multiplied by an  $E_\gamma$-dependent factor.
The fractions of the $L_2=0$, 1, and 2 contributions
are limited to
38.6\%--49.3\% (49.1\%--57.5\%),
50.0\%--61.4\% (41.3\%--50.9\%), and
0.0\%--2.0\% (0.0\%--2.3\%)
for $E_\gamma= 1.01$--$1.15$ ($0.95$--$1.01$) GeV, respectively,
to match the results from 
the mass distribution analysis,
giving a minimum $\chi^2$ of 43.3.
The solid curves (magenta) in Fig.~\ref{fig3}
show the angular distributions calculated for the best-fit solution.
The dashed (blue), dot-dashed (red), and two-dot-dashed (cyan) curves show 
the $L_2=0$, 1, and 2 contributions, respectively.
The long-dashed curves (black) represent the interference effects;
those between even and odd $L_2$s (between $L_2=0$ and 2) are observed in the $\pi^0$ ($\eta$) angular distributions.
The $L_2=1$ amplitudes and the $L_2=0$ and 2 interference
make the $\eta$ angular distribution convex upward.
Regarding $\pi^0\mathcal{D}_{\eta d}$, the major component is $0^-$
($\sim 47\%$), 
and the amplitudes are distributed widely to other 
 $1^+$ and $2^\pm$ components.
For $\eta\mathcal{D}_{12}$,
the major component is $2^+$ ($\sim 100\%$).

We also estimate the excitation function for the events with  $M_{\eta d}<2.47$ GeV,
as represented by the squares (cyan) in Fig.~\ref{fig1}.
It forms a bump
at $\sim 1$ GeV corresponding to the $\gamma d$-CM energy 
of $\sim 2.69$ GeV. 
The observed broad bump corresponds to some resonances
because the expected excitation function 
monotonically increases for the three-body phase-space contribution with  $M_{\eta d}<2.47$ GeV as plotted by the dotted curve (green) in Fig.~\ref{fig1}.
Loosely-coupled isovector $S$-wave molecules
$N$-$\Delta(1620)1/2^-$ and $N$-$N(1650)1/2^-$
would play the role of a doorway to the $\pi^0 \mathcal{D}_{\eta d}$ system
with $0^-$.
It should be noted that 
neither  $\Delta(1620)1/2^-$ nor $N(1650)1/2^-$
is considered a contributor to the elementary
$\gamma N\to \pi^0\eta N$ reaction
(the main contributor is $\Delta(1700)3/2^-$)~\cite{doer06,ajak08,fix3,mainz2,gutz14}.
In contrast,
$N$-$N(1720)3/2^+$ is a candidate doorway to $\eta\mathcal{D}_{12}$
with $2^+$
although the branching ratio of $N(1720)3/2^+ \to \eta N $ 
is only a few percent~\cite{pdg}.

In summary,
the $\eta d$ threshold structure has been experimentally studied 
in the $\gamma{d}${$\to$}$\pi^0\eta{d}$ reaction
at $E_\gamma<1.15$ GeV.
It is found that the $M_{\eta d}$ dependence of $d\sigma/dM_{\eta d}$ is 
quite different from the behavior of the three-body phase space
but shows a strong enhancement near the threshold,
which changes in shape depending on the incident energy.
An analysis incorporating the known resonance $\mathcal{D}_{12}$
in the $\pi^0d$ channel has revealed the existence
of a narrow resonance-like structure in the $S$-wave
$\eta d$ system, $\mathcal{D}_{\eta d}$.
Applying a Flatt\'e parameterization to $\mathcal{D}_{\eta d}$,
we have obtained the mass 
of $2.427_{-0.006}^{+0.013}$ GeV and
the width $\Gamma=\Gamma_0+g p_\eta c$ with 
$\Gamma_0=0.029_{-0.029}^{+0.006}$ GeV and $g=0.00_{-0.00}^{+0.41}$,
where $p_\eta$ denotes the $\eta$ momentum in the $\eta d$-CM frame,
and $g$ denotes the effective coupling to the $\eta d$ channel.
The $S$-wave resonance states with widths broader than 0.05 GeV
are ruled out.
The $\mathcal{D}_{\eta d}$ system
would be the predicted $\eta d$ bound state, or
an $\eta d$ virtual state originating from 
strong $\eta d$ attraction.
The major component of the $\pi^0 \mathcal{D}_{\eta d}$ system
is found to be $0^-$ 
from the $\pi^0$ and $\eta$ angular distributions for 
the events with $M_{\eta d}<2.47$ GeV.

\begin{acknowledgments}
The authors express gratitude to the ELPH accelerator staff for stable operation 
of the accelerators in the FOREST experiments.
They acknowledge Mr.\ Kazue~Matsuda, Mr.~Ken'ichi~Nanbu, and Mr.~Ikuro~Nagasawa for their technical assistance in the FOREST experiments.
They also thank Prof.\ Alexander I. Fix and Prof.\ Mikhail Egorov 
for their theoretical calculations and fruitful discussions.
They are grateful for valuable discussions with Prof.\ Heinz A Clement, 
Prof.\ Atsushi Hosaka, Prof.\ Kiyoshi Tanida, and Prof.\ Hiroyuki Fujioka.
They acknowledge Prof.\ Bernd Krusche for the numerical values of the total cross sections 
measured at the Mainz MAMI facility.
This work was supported in part by the Ministry of Education, Culture, Sports, Science 
and Technology, Japan (MEXT) and the
Japan Society for the Promotion of Science (JSPS)
through Grants-in-Aid 
for Specially Promoted Research No.\ 19002003,
for Scientific Research (A) Nos.\ 24244022 and 16H02188,
for Scientific Research (B) Nos.\ 17340063 and 19H01902,
for Scientific Research (C) No.\ 26400287, and
for Scientific Research on Innovative Areas Nos.\ 19H05141 and 19H05181.
\end{acknowledgments}

\end{document}